\documentclass[a4paper, amsfonts, amssymb, amsmath, reprint, showkeys, nofootinbib, twoside]{revtex4-1}
\usepackage[english]{babel}
\usepackage[utf8]{inputenc}
\usepackage[colorinlistoftodos, color=green!40, prependcaption]{todonotes}
\usepackage{amsthm}
\usepackage{mathtools}
\usepackage{physics}
\usepackage{xcolor}
\usepackage{graphicx}
\usepackage[left=23mm,right=13mm,top=35mm,columnsep=15pt]{geometry} 
\usepackage{adjustbox}
\usepackage{placeins}
\usepackage[T1]{fontenc}
\usepackage{lipsum}
\usepackage{csquotes}
\usepackage[pdftex, pdftitle={Article}, pdfauthor={Author}]{hyperref} 
\begin{document}
\title{Charged Higgs Phenomenology in di-bjet channel with $H^{\pm} \rightarrow W^{\pm}h$ in 2HDM Type-II using Machine Learning Technique}

\author{Kanhaiya Gupta}
    \email[Correspondence email address: ]{kanhaiya.gupta@uni-bonn.de}
    \affiliation{Physikalisches Institut, Universität Bonn, Nussallee 12, 53115 Bonn, Germany}

\date{\today} 

\begin{abstract}
The latest LHC collaborations results on $\sigma_{H^{\pm}}BR(H^{\pm} \rightarrow \tau^{\pm}\nu)$ and $\sigma_{H^{\pm}}BR(H^{+} \rightarrow t\Bar{b})$ are used to
impose constraints on the charged Higgs $H^{\pm}$ parameters
within the Two Higgs Doublet Model (2HDM). But it leaves $1.5 \leq tan \beta \leq 3$ window unexplored where the $BR(H^{\pm} \rightarrow W^{\pm}h)$ becomes sizable for $m_{H^{\pm}} > m_{t}$. In this manuscript $H^{\pm} \rightarrow W^{\pm}h$ is investigated with neutral Higgs boson decaying to a pair of b-quarks and the discovery prospects of charged Higgs boson is discussed. In particular, the analysis is optimized by putting the kinematic cuts and prospects of using Machine Learning Technique to derive values of $\sigma\times BR$ needed for a $5 \sigma$ discovery at the LHC.
\end{abstract}


\maketitle

\section{Introduction}  \label{sec:introduction}

The Standard Model (SM) of particle physics describes with high accuracy a multitude of experimental and observational data ranging ranging from the smallest energy scales up to the scale of several TeV set by the center-of-mass energy of the Large Hadron Collider (LHC) at CERN. With the recent discovery of the Higgs boson\cite{aad2012observation, chatrchyan2012observation}, the particle content of the SM corresponds to a renormalizable theory and is in this sense complete. Nevertheless, the SM does not accommodate phenomena such as gravity or dark matter and dark energy inferred from cosmological observations, prompting theoretical work on its extensions. Over the years, many models addressing these shortcomings purport to go "beyond the Standdard Model (BSM)". These models predict new particles in the form of new heavy vector, scalar or fermion resonances which can potentilly be observed at the CERN LHC. 

On the earliest BSM scenarios to emerge was Technicolor (TC)\cite{hill1993spontaneously} but these did not rely on the existence of fundamental scalar particle (Higgs). With the discovery of the Higgs boson at CERN LHC implies the Higgsful extension of the SM. One of the simplest extensions of the Standard Model (SM) Higgs sector is the Two-Higgs-Doublet Model (2HDM). A recent measurements from the B-Factory experiments BABAR, Belle and LHCb have reported large disagreements (at the $3.9 \sigma$ level) in semi-tauonic decays ($B \rightarrow \tau \nu$) involving ratios of charmed final states\cite{Lees_2012,Huschle_2015, Aaij_2015, Proceedings:2015ona}.  This disagreement is an interesting anomaly that deserves more experimental and theoretical study, in particular due to the absence of any clear new physics signal from the experiments at the energy frontier of particle physics. One of the possible explanation is the existence of BSM charged Higgs ($H^{\pm}$) can enhance this particular decay.

This manuscript is organized as follows: Section \ref{sec:2hdm} presents the necessary details of the model under investigation. Section \ref{sec:cross_section_decay} discusses the production mechanism and the decays of the charged Higgs in this model. Section \ref{sec:expt_limit} presents the current bounds and in Section \ref{sec:search_lhc}, the discovery prospects of Charged Higgs is presented based on kinematic cuts while Section \ref{sec:machine_learn} presents the setup for Machine Learning technique. The manuscript concludes in Section \ref{sec:conclusion} with a summary of the key results.
\section{Two-Higgs-Doublet Model}  \label{sec:2hdm}
This section presents an overview of the two-Higgs-doublet model (2HDM) that will be used as an operating example of our phenomenological model. The discussion here is very brief, only touching the parts relevant to this manuscript. A comprehensive review of this model is given in ref\cite{Branco_2012}. The scalar potential for two doublets $\Phi_{1}$ and $\Phi_{2}$ with hypercharge $+1$ is 
\begin{equation}
\begin{split}
    V(\Phi_{1},\Phi_{2}) &= m_{1}^{2}\Phi_{1}^{\dagger}\Phi_{1} + m_{2}^{2}\Phi_{2}^{\dagger}\Phi_{2} - (m_{12}^{2}\Phi_{1}^{\dagger}\Phi_{2} + h.c) \\ &+ \frac{1}{2}\lambda_{1}(\Phi_{1}^{\dagger}\Phi_{1})^{2} + \frac{1}{2}\lambda_{2}(\Phi_{2}^{\dagger}\Phi_{2})^{2} + \lambda_{3}(\Phi_{1}^{\dagger}\Phi_{1})(\Phi_{2}^{\dagger}\Phi_{2}) \\ &+ \lambda_{4}(\Phi_{1}^{\dagger}\Phi_{2})(\Phi_{1}^{\dagger}\Phi_{2}) + [\frac{1}{2}\lambda_{5}(\Phi_{1}^{\dagger}\Phi_{2})^{2} + h.c ]
\end{split}
\end{equation}
where all the parameters are real for simplicity. The complex scalar doublets $\Phi_{i} (i=1,2)$ can be parametrized as

\begin{equation}
    \Phi_{i}(x) =  \begin{pmatrix}
\phi^{+}(x) \\
\frac{1}{\sqrt{2}}(v_{i} + \rho_{i}(x) + \iota \eta_{i}(x))
\end{pmatrix}
\end{equation}
with $v_{1}, v_{2} \geq 0$  being VEVs satisfying $v=\sqrt{v_{1}^{2} + v_{2}^{2}} = 246.22$ GeV.
After electroweak symmetry breaking (EWSB), three of the eight degrees of freedom in the Higgs sector of the 2HDM are eaten by the Goldstone boson which give mass to $W^{\pm}$ and Z. The remaining five becomes physical Higgs boson. After the minimization of the potential, the 2HDM has seven independent parameters.

  $m_{h}$, $m_{H}$, $m_{A}$, $m_{H^{\pm}}$, $\alpha$, $tan\beta$, $m_{12}^{2}$

where $tan\beta = \frac{v_{1}}{v_{2}}$ and $\alpha$ is the mixing angle. Of the particular interest to this manuscript is $m_{H^{\pm}}$. 

\section{Collider Phenomenology} \label{sec:collider_phenomenology}

\subsection{Cross sections, decay rates and search strategy}

\label{sec:cross_section_decay}
The feasible study of $H^{\pm} \rightarrow hW^{\pm}$ is begun by setting up calculations and specific search strategy is employed. Putting constraint on $sin(\beta - \alpha)$ automatically implies the model dependent search but the analysis will be presented in a more general framework without recourse to the specific model discussed in section 2. The first step in the analysis is the identification of the production and decay modes. The dominant production process at the LHC for a $H^{\pm}$ heavier than the top quark is its associated production with a single top, with the relevant sub-processes being $gb\rightarrow tH^{-}$ and $gg\rightarrow t\Bar{b}H^{-}$ (plus charge comgugated channel). The spin/color summed/averaged squared amplitude for the $gb\rightarrow tH^{-}$ production process is given ref\cite{Moretti_2002}
\begin{equation}
\begin{split}
 \vspace{-2.0cm} \overline{\abs{\iota \mathcal{M}}^{2}} = \frac{g_{qH^{\pm}}^{2}}{2m_{W}^{2}}\frac{g_{s}^{2}g_{2}^{2}}{4N_{c}}\abs{V_{tb}}^{2} \frac{(u - m_{H^{\pm}}^{2})^{2}}{s(m_{t}^{2} - t)} \cross \\ \left[  1 + 2 \frac{m_{H^{\pm}}^{2} - m_{t}^{2}}{u - m_{H^{\pm}}^{2}}  \cross \left( 1 + \frac{m_{t}^{2}}{t - m_{t}^{2}} + \frac{m_{H^{\pm}}^{2}}{u - m_{H^{\pm}}^{2}} \right) \right]
\end{split}
\end{equation}

where $g_{s}$ and $g_{2}$ are the $SU(3)_{C}$ and $SU(2)_{C}$ gauge couplings, $N_{c} = 3$ is the number of colours and $V_{tb}$ is the relevant CKM matrix element. The coupling $g_{qH^{\pm}}$ is the model dependent parameter. The expression for $g_{qH^{\pm}}^{2}$ in the different 2HDM are in Table-I\cite{enberg2015charged}. 

\begin{table}
\caption{The expressions for $g^{2}_{qH^{\pm}}$ in the 2HDMs Type-I \& II considered in this paper.}
\begin{tabular*}{\columnwidth}{@{\extracolsep{\fill}}lll@{}}
\hline\noalign{\smallskip}
 & 2HDM-I & 2HDM-II  \\
 \hline  \\
 $g^{2}_{qH^{\pm}}$ &  $m_{b}^{2}cot^{2}\beta + m_{t}^{2}cot^{2}\beta$  & $m_{b}^{2}tan^{2}\beta + m_{t}^{2}cot^{2}\beta$ \\
\noalign{\smallskip}\hline 
\end{tabular*}
\end{table}

\begin{figure}[ht]
    \centering
    \includegraphics[width=0.5\textwidth]{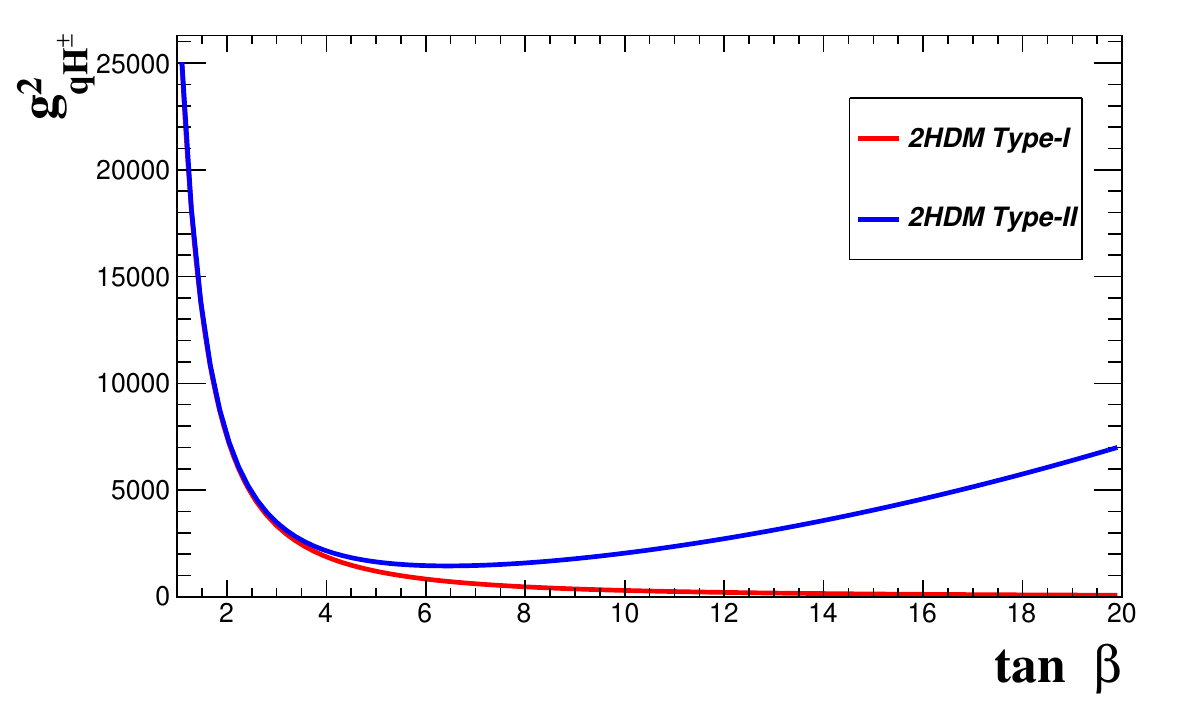}
    \caption{Coupling $g^{2}_{qH^{\pm}}$ vs tan $\beta$ for Type-I and Type-II 2HDM}
\end{figure}

From Fig. 1, the amplitude for 2HDM Type-II for the $gb \rightarrow tH^{-}$ process is maximal for either small or large $tan\beta$. Fig. 2 show the phase-space between the minimum value of the charged Higgs and $tan\beta$. For both Type-I \& Type-II, $tan\beta \leq 2$ is the feasible region for the minimum value of the charged Higgs boson. While Fig. 3 displays the experiments constraints on tan$\beta$ vs $m_{H^{\pm}}$ proposed by the Belle Detector. The decay however can proceed via multiple mechanism depending on the strength of couplings in the particular model, and availability of phase space. The light charged Higgs boson would decay almost exclusively into a (hadronic or leptonic) $\tau$ lepton and its associated neutrino for $tan\beta \geq 1$. When the top-bottom channel is kinematically open, then $H^{\pm} \rightarrow t\Bar{b}$ would compete with $H^{\pm} \rightarrow hW^{\pm}$, $HW^{\pm}$, $AW^{\pm}$ decay as well various SUSY channel in the MSSM. At the tree level, the decay width of $H^{\pm} \rightarrow w^{\pm}h$ is given by equation 4. The strength of $hW^{\pm}$ channel is proportional to $cos^{2}(\beta - \alpha)$ and it is therefore absent for $sin(\beta - \alpha) = 1$. 
\begin{equation}
\begin{split}
\Gamma(H^{\pm} \rightarrow W^{\pm}h) &= \frac{\lambda^{3/2}(m^{2}_{H^{\pm}}m_{W}^{2},m_{h}^{2})}{64\pi m_{H^{\pm}}^{3}m_{W}^{2}}\frac{e^{2}}{sin^{2}{\theta_{W}}}cos^{2}(\beta - \alpha) \\
\lambda(x,y,z) &= x^{2} + y^{2} + z^{2} - 2xy - 2xz - 2yz
\end{split}
\end{equation}

\begin{figure}[ht]
    \centering
    \includegraphics[width=0.48\textwidth]{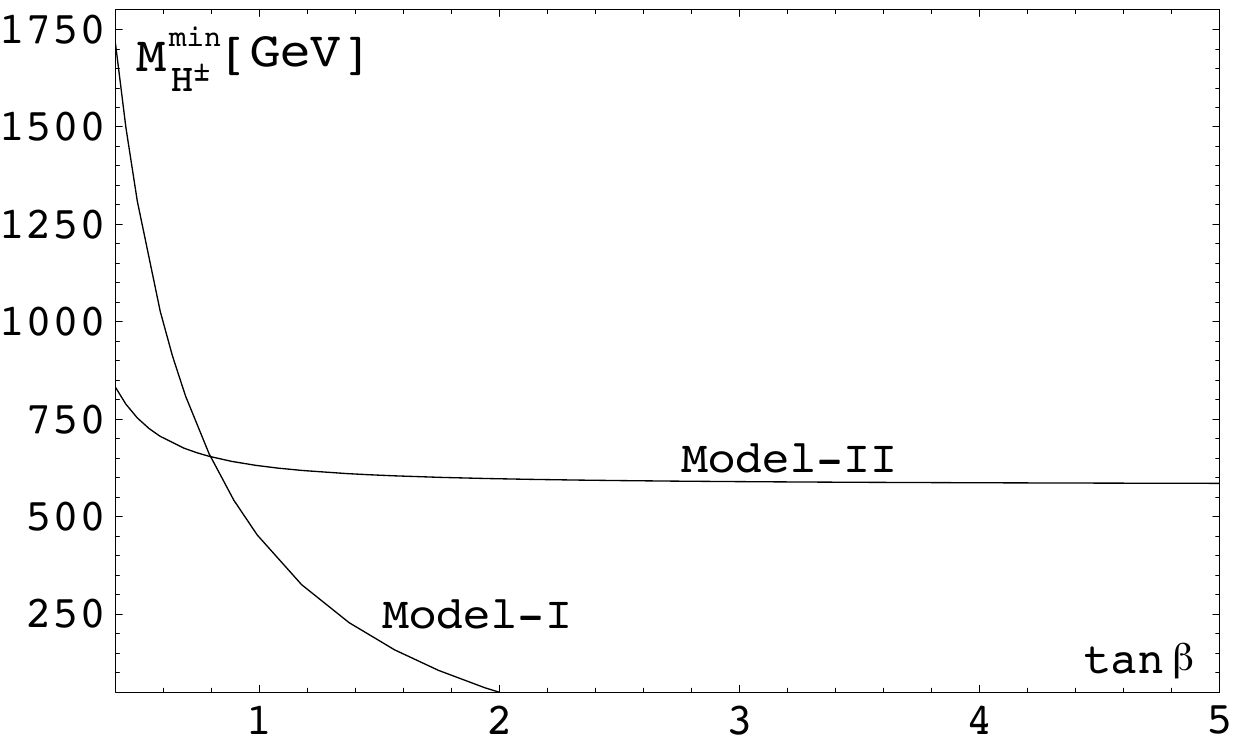}
    \caption{95\% C.L. lower bounds on $M_{H^{\pm}}$ as functions of $tan \beta$. Above $tan\beta \approx$ 2, the Model I bound becomes weaker than the LEP one ($\approx$ 80 GeV), while the Model II one gets saturated by its $tan\beta \rightarrow \infty$ limit ($\approx$ 580 GeV).\cite{misiak2017weak}}
\end{figure}

\begin{figure}[ht]
\centering
\includegraphics[width=0.5\textwidth]{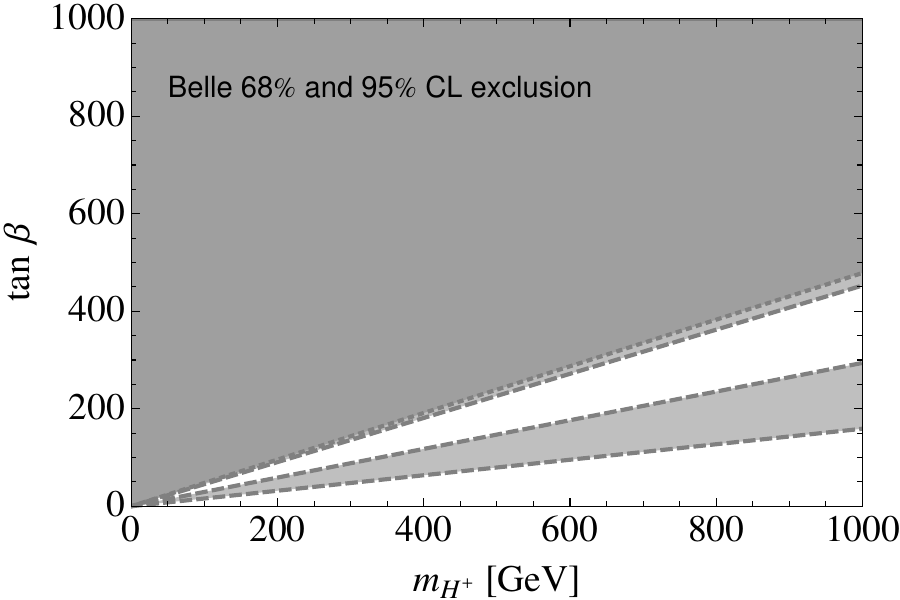}
\caption{Region of the $tan\beta$-$m_{H^{\pm}}$ excluded at the different confidence interval (CI) by the Belle experiment. The 68$\%$ CI is shown in light grey whereas the dark grey shows 95$\%$ CI \cite{hamer2016search}.}
\end{figure}

 The further constraints on $cos(\beta - \alpha)$ for both 2HDM Type-I and 2HDM Type-II are given in (a) and (b) of Fig. 5 in ref\cite{constra}. Based on recent LHC and Belle collaborations results discussed in Section 3.1, we choose tan$\beta$ = 1.3 and sin$(\beta - \alpha)$ = 0.7 for the phenomenological study in the 2HDM Type-II model. A similar study of the allowed parameter space of the 2HDM Type II is also performed in the ref\cite{atkinson}. In addition to constraining the parameter spaces of the new physics models, knowledge of the mass of observed Higgs also provides an additional handle in identifying the $H^{\pm} \rightarrow W^{\pm}h$ decay. In this manuscript the $h \rightarrow b \Bar{b}$ is focused as it generally has a substantial branching ratio (BR) and allows for a full reconstruction of Higgs boson. 

\begin{figure}
\centering
\includegraphics[width=0.5\textwidth]{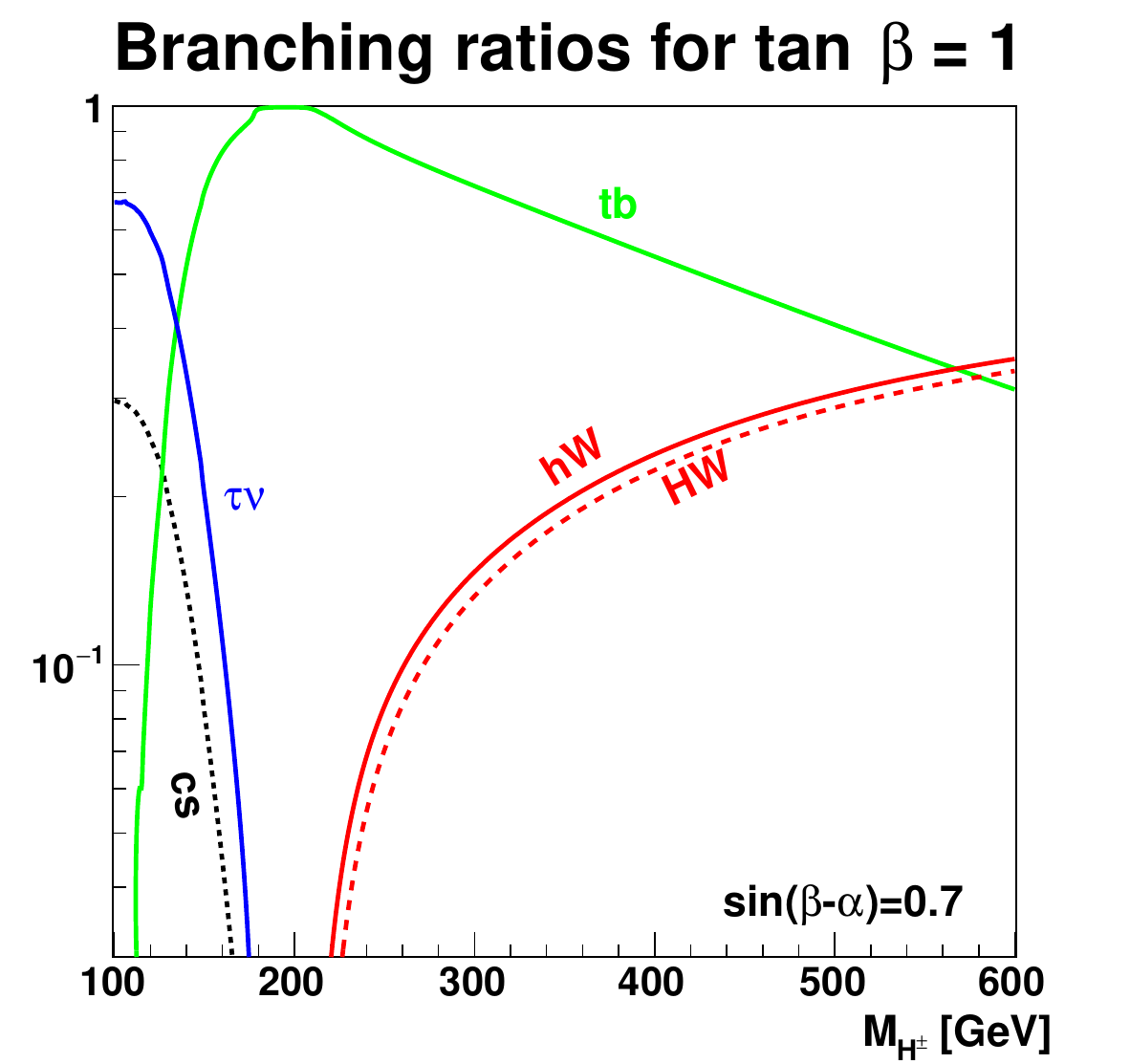}
\caption{Charged-Higgs branching ratios vs. $M_{H^{\pm}}$ , for tan$\beta$ = 1 and sin $(\beta - \alpha)$ = 0.7. Here, two light neutral Higgs bosons h and H (125 and 130 GeV) are considered \cite{akeroyd2017prospects}}
\end{figure}

\subsection{Experimental limits}
\label{sec:expt_limit}
The LEP experiments \cite{collaborations2013search} have given limits on the mass of the charged Higgs boson in 2HDM from the charged Higgs searches in Drell–Yan events, $e^{+}e^{-} \rightarrow Z/\gamma \rightarrow H^{+}H^{-}$, excluding $m_{H^{\pm}} \leq 80$ GeV (Type II) and $m_{H^{\pm}} \leq 72.5$ GeV (Type I) at 95\% confidence level. The Heavy Flavor Averaging Group (HFAG)\cite{heavyflavoraveraginggroup2014averages} the $B \rightarrow Xs \gamma$ decay puts a very strong constraint on Type II and Type Y 2HDM, excluding $m_{H^{\pm}} \leq 580$ GeV. At the LHC several searches have been carried out for $H^{\pm}$'s lighter as well as heavier than the top quark. The CMS collaboration has recently released exclusion limits \cite{cms2013search} for a $H^{\pm}$ lying in the  mass range (180-600) GeV in $gg \rightarrow tH^{-}\Bar{b}$ production modes and in $H^{\pm} \rightarrow tb$ and $H^{\pm} \rightarrow \tau \nu$ decay modes based on 19.7 $fb^{-1}$ of data collected at $\sqrt{s} = 8$ TeV. The analysis \cite{aad2015search} based on same dataset via $t\Bar{t} \rightarrow H^{\pm}W^{\pm}b \Bar{b}$ production mode and in the $H^{\pm} \rightarrow \tau \nu$ channel provides an exclusion limit of 80 GeV $\leq m_{H^{\pm}} \leq$ 160 GeV.  The ATLAS Collaborations \cite{aad2016search} based on 19.5 $fb^{-1}$ dataset at $\sqrt{s} = 8$ TeV for the same production and decay modes provides an exclusion limits for 80 GeV $\leq m_{H^{\pm}} \leq$ 160 GeV and 180 GeV $\leq m_{H^{\pm}} \leq$ 1 TeV. 

\begin{figure*}
\begin{minipage}[t]{.49\textwidth}
\centering
\includegraphics[width=\textwidth]{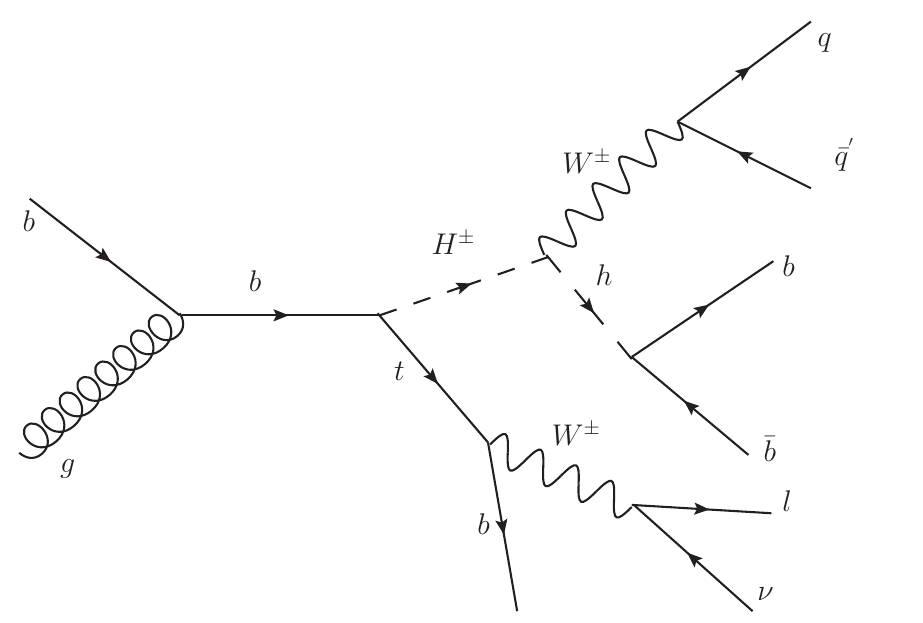}
\caption{Feynman diagram showing the dominant production of Charged Higgs from $gb \rightarrow tH^{\pm}$ and $jjbb\Bar{b} l \nu$ final states. The Feynman diagram is drawn using JaxoDraw software\cite{binosi2004jaxodraw}.}
\end{minipage}
\hfill
\begin{minipage}[t]{.49\textwidth}
\centering
\includegraphics[width=\textwidth]{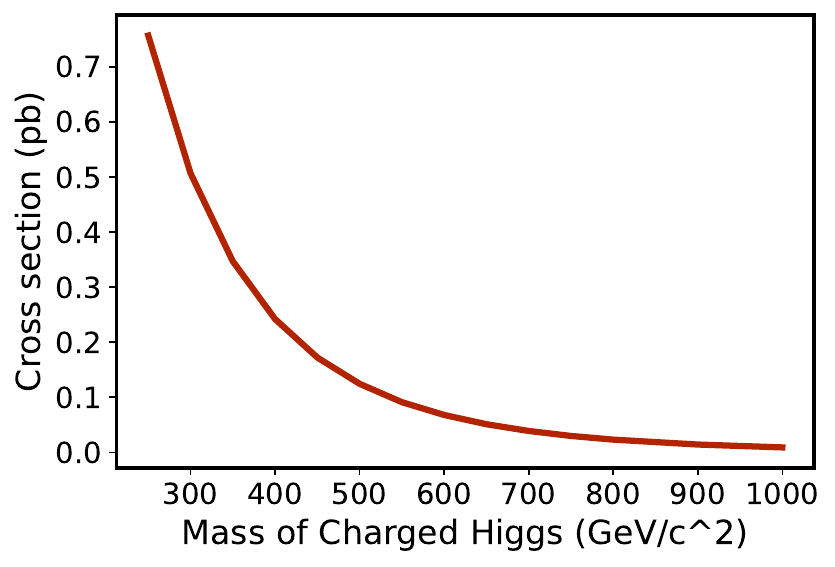}
\caption{Charged-Higgs boson production cross-section with tan$\beta$ = 1 and sin $(\beta - \alpha)$ = 0.7. Here, two light neutral Higgs bosons h and H (125 and 130 GeV) are considered}
\end{minipage}
\end{figure*}

The two dominant decay channel, tb and $\tau \nu$ leave $1.5 \leq tan \beta \leq 3$ window unexplored for $m_{H^{\pm}} > m_{t}$. For the smaller values of tan$\beta$, the $BR(H^{\pm} \rightarrow W^{\pm}h)$ becomes sizable as shown in  Fig. 6 for tan$\beta$ = 1 and sin$(\beta - \alpha)$ = 0.7. Here, two light neutral Higgs bosons h and H (125 and 130 GeV) are considered.

\begin{figure*}
\begin{minipage}[t]{.49\textwidth}
\centering
\includegraphics[width=\textwidth]{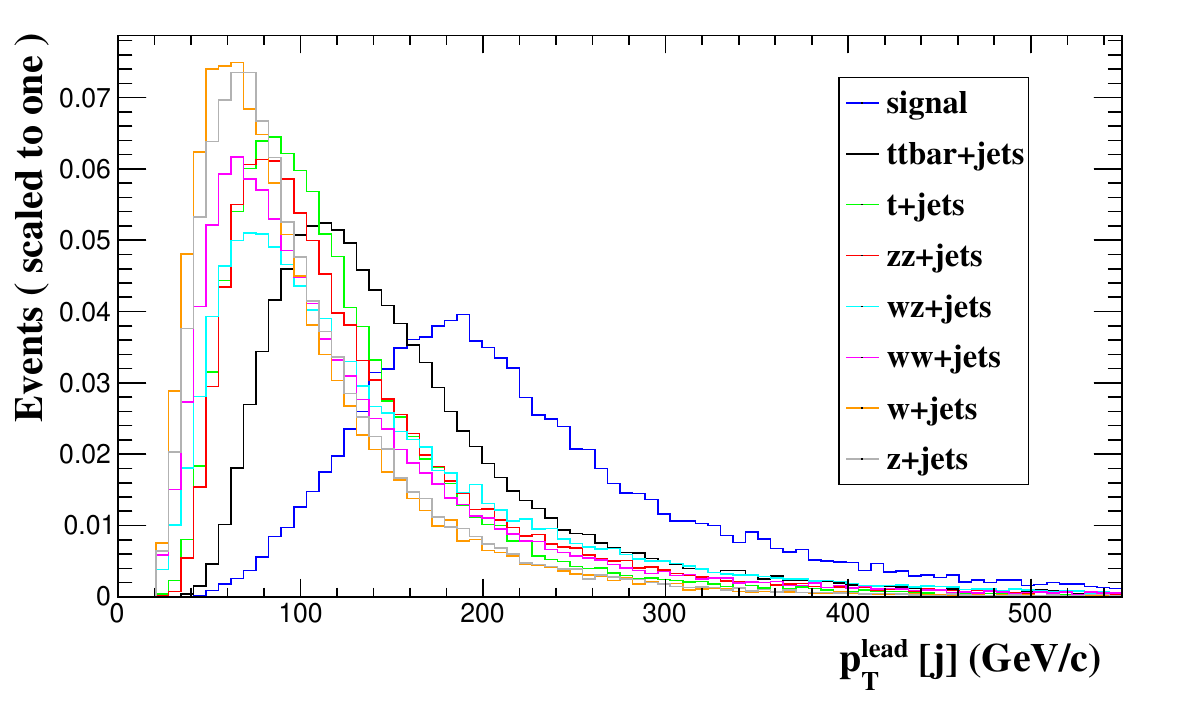}
\caption{Transverse momentum of the leading jet for both signal and background.}
\end{minipage}
\hfill
\begin{minipage}[t]{.49\textwidth}
\centering
\includegraphics[width=\textwidth]{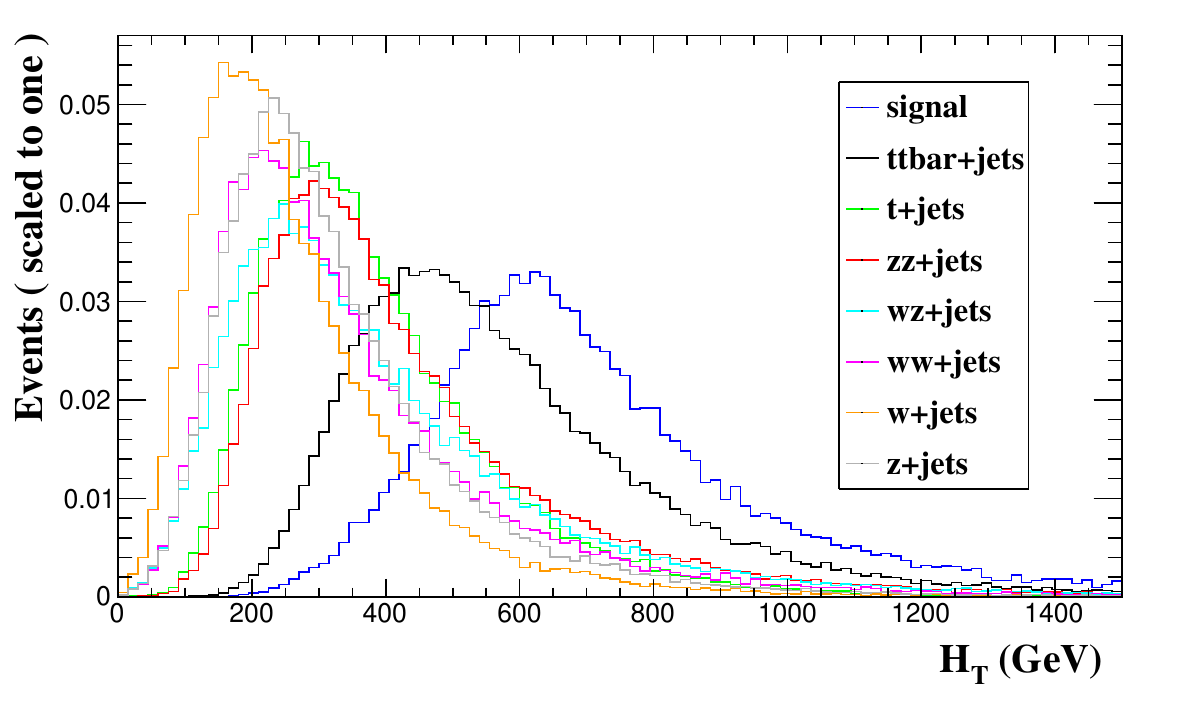}
\caption{Transverse hadronic energy $H_{T}$ distribution for both signal and background.}
\end{minipage}
\begin{minipage}[t]{.49\textwidth}
\centering
\includegraphics[width=\textwidth]{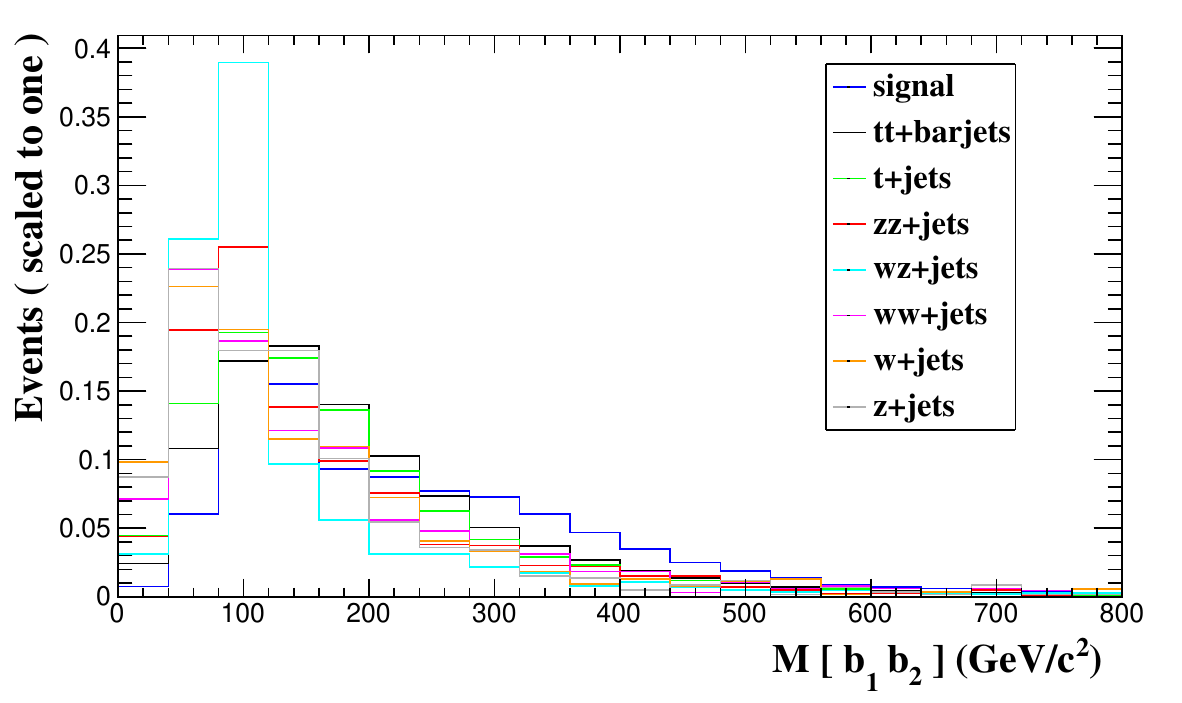}
\caption{ The invariant mass distribution $M_{bb}$ for both the signal (corresponding to $M_{h} = 125$ GeV) and the
background. }
\end{minipage}
\hfill
\begin{minipage}[t]{.49\textwidth}
\centering
\includegraphics[width=\textwidth]{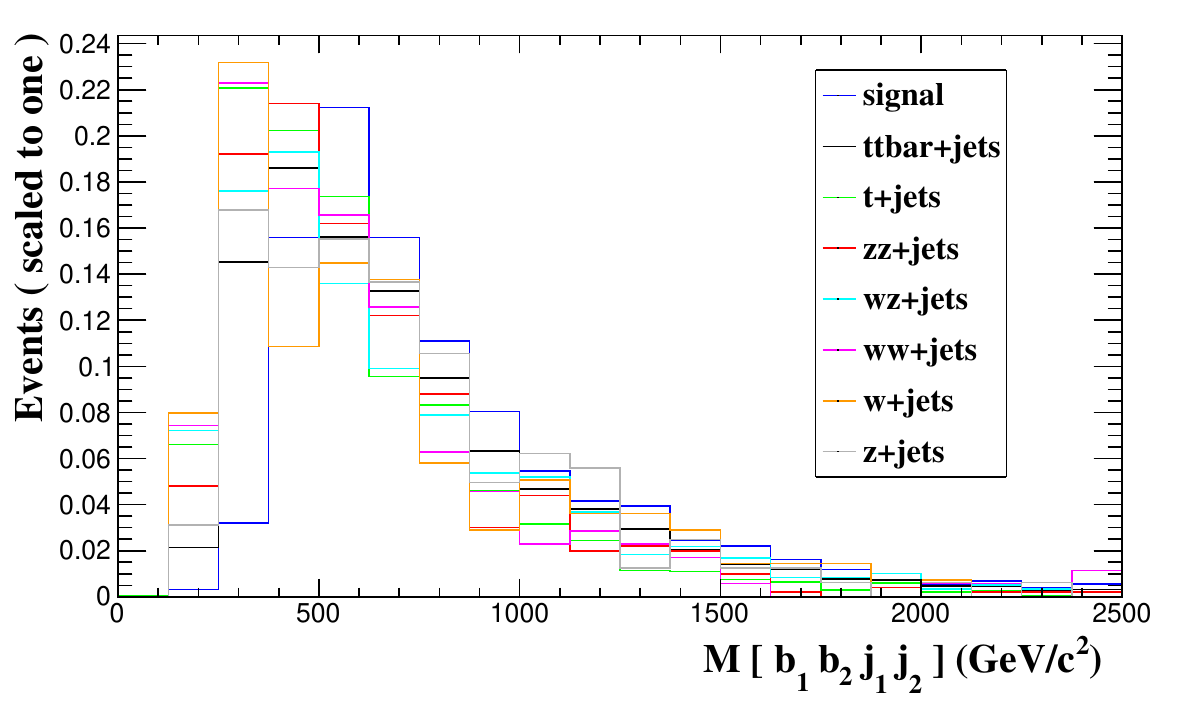}
\caption{ The invariant mass distribution $M_{bbjj}$ for both the signal (corresponding to  $M_{H^{+}} = 500$ GeV) and the
background. }
\end{minipage}
\end{figure*}

\begin{table*}[ht]
\caption{Cut flow chart showing the number of events after passing the cut selection}
\begin{tabular*}{\textwidth}{@{\extracolsep{\fill}}lrrrrrrrrr@{}}
\hline\noalign{\smallskip}
Cut selection & Signal & ttbar+jets & t+jets  & ZZ+jets & WZ+jets &  WW+jets & Z+jets & W+jets & $\frac{S}{\sqrt{S+B}}$ \\ \hline 
Initial & 30000 & 70000 & 50000 & 50000 & 50000 & 50000 & 50000 & 50000  & 47.43 \\
3 $\leq N_{jets} \leq$ 10 & 29588 & 67674 & 48546 & 49432 & 43670 & 42655 & 42910 & 41332 & 48.92 \\
 $ N_{bjets} \geq$ 3 & 9729 & 5250 & 1207 & 131 & 224 & 52 & 31 & 34  & 75.38 \\
$ MET \geq 20$ GeV & 9138 & 4857 & 1110 & 70 & 204 & 46 & 21 & 29 & 73.46 \\
 $H_{T} \geq 250$ GeV & 9132 & 4833 & 1048 & 69 & 187 & 44 & 19 & 24 & 73.69\\
\hline\noalign{\smallskip}
\end{tabular*}
\end{table*}

\begin{figure*}[ht]
\begin{minipage}[t]{.49\textwidth}
\centering
\includegraphics[width=\textwidth]{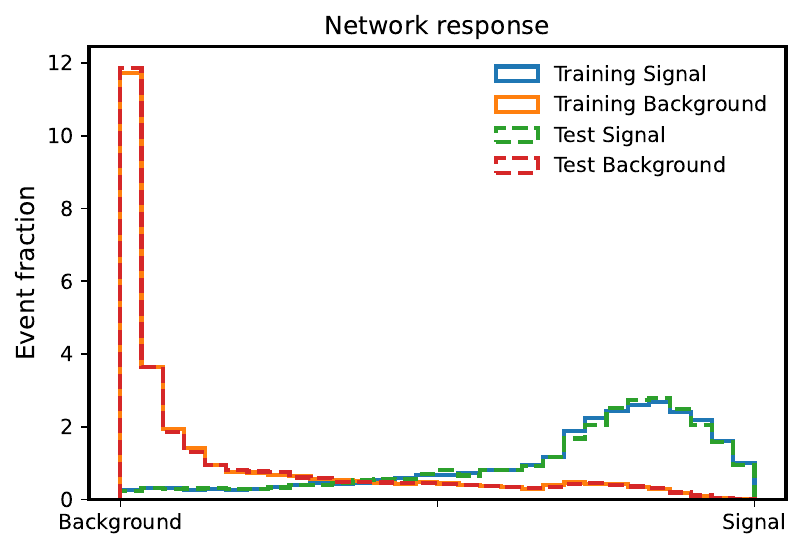}
\caption{Neural network response for the training and test samples for signal against t+jets background}
\end{minipage}
\hfill
\begin{minipage}[t]{.49\textwidth}
\centering
\includegraphics[width=\textwidth]{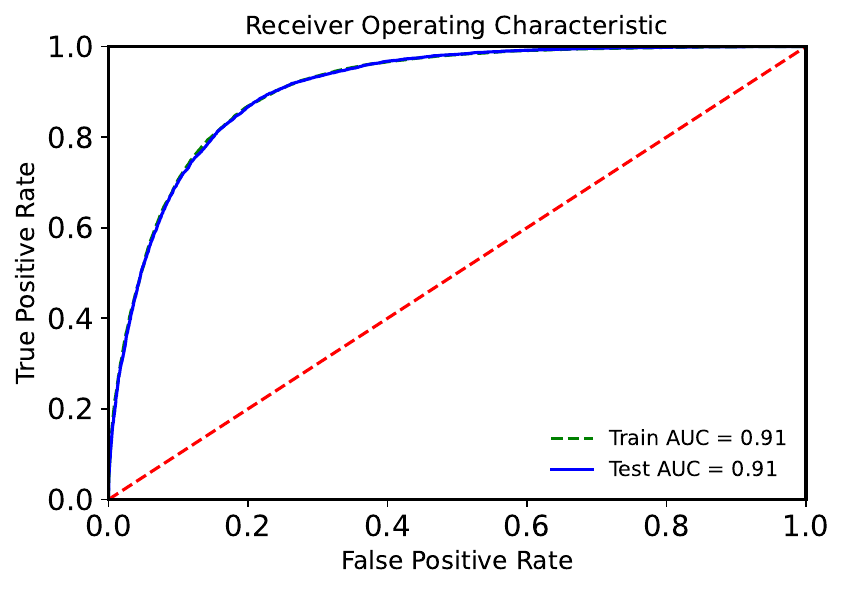}
\caption{ROC curve for the training and test samples for signal against t+jets background}
\end{minipage}
\begin{minipage}[t]{.49\textwidth}
\centering
\includegraphics[width=\textwidth]{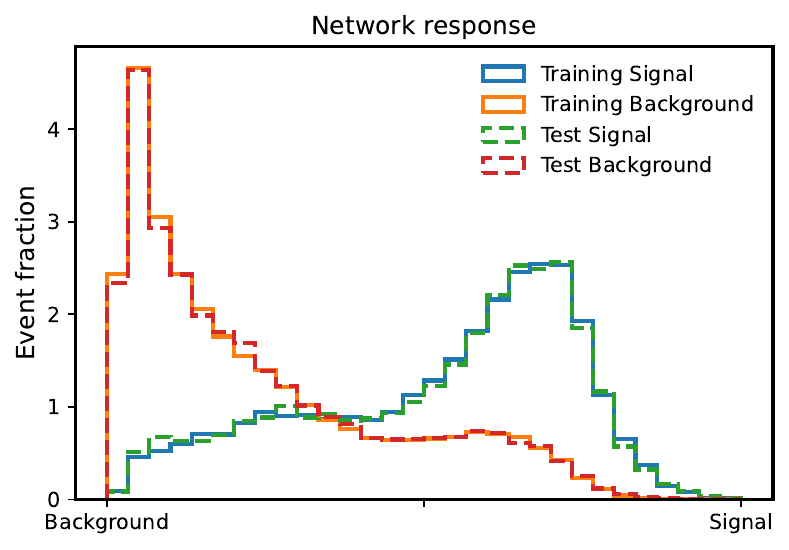}
\caption{Neural network response for the training and test samples for signal against ttbar+jets background}
\end{minipage}
\hfill
\begin{minipage}[t]{.49\textwidth}
\centering
\includegraphics[width=\textwidth]{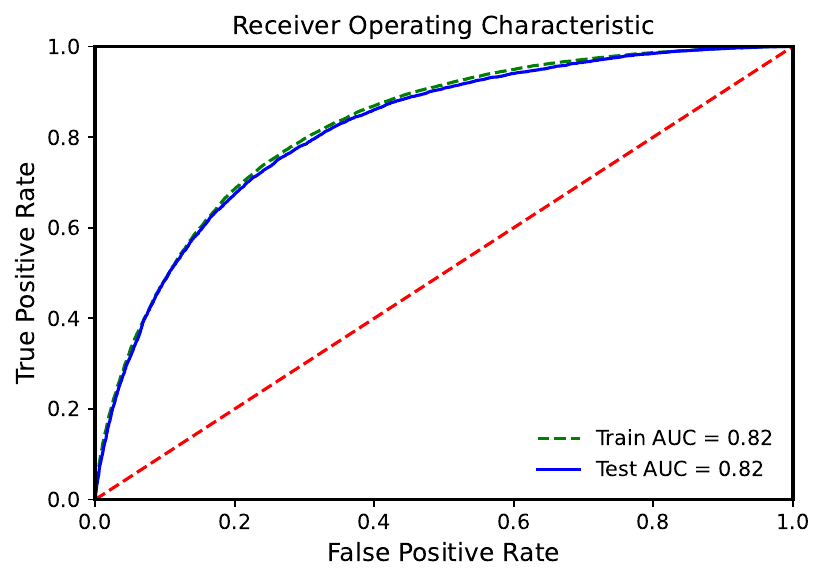}
\caption{ROC curve for the training and test samples for signal against ttbar+jets background}
\end{minipage}
\end{figure*}

\begin{figure*}
\begin{minipage}[t]{.49\textwidth}
\centering
\includegraphics[width=\textwidth]{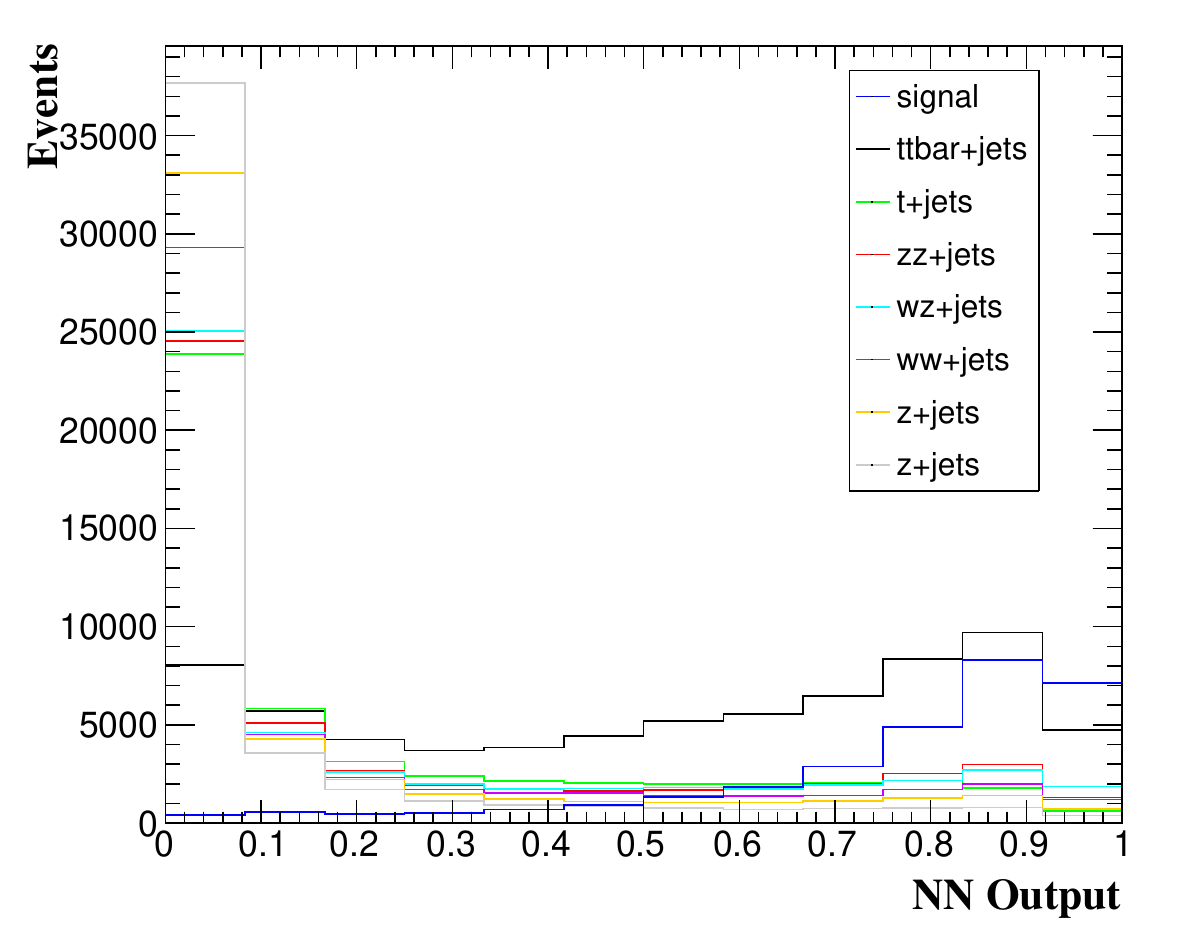}
\caption{Neural network prediction of all the background besides ttbar+jets for the training of signal against t+jet. }
\end{minipage}
\hfill
\begin{minipage}[t]{.49\textwidth}
\centering
\includegraphics[width=\textwidth]{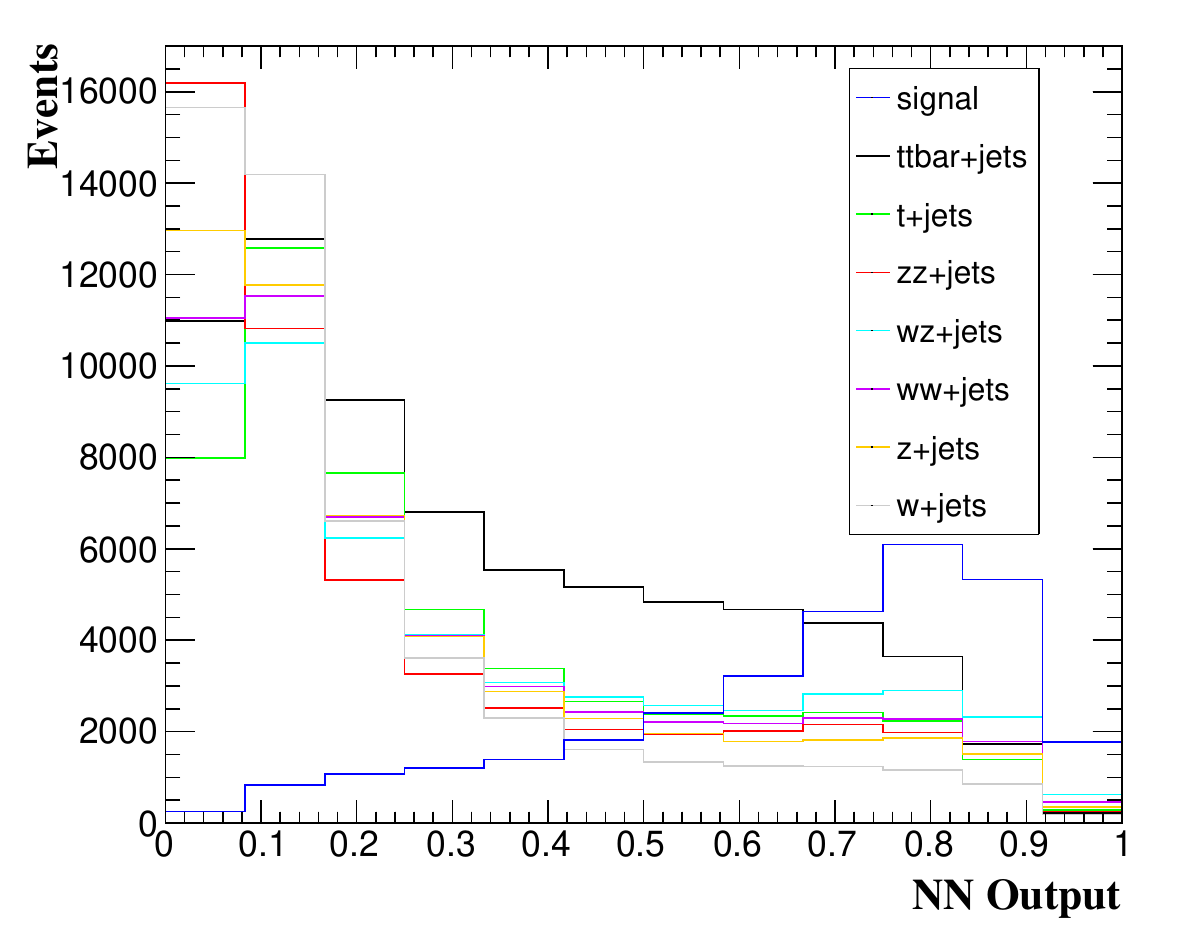}
\caption{Neural network prediction of all the background for the training of signal against ttbar+jet.}
\end{minipage}
\end{figure*}

\subsection{Search prospects at LHC}
\label{sec:search_lhc}
After setting the required input parameters, the complete channel chosen to study the prospects of discovery is $p p \rightarrow H^{\pm}t  \rightarrow j j l \nu b b b$. where $H^{\pm} \rightarrow W^{\pm}h$. The charged Higgs boson is produced via $gb \rightarrow H^{\pm}t$. The process $h \rightarrow b b$ has the largest branching ratio which is nearly 0.57\% but suffers from high backgrounds from QCD multi-jet. On the other hand, the leptonic decay of W boson lowers the background. The Feynman diagram corresponding to the dominant production process and the final decay channels used in our analysis is shown in Fig. 5. The production cross section depends on the coupling $g_{qH^{\pm}}$, to get an idea of the numbers involved, Fig. 4 shows the production cross-section for the benchmarks region 300-800 GeV. The data simulation is performed using the MadGraph5$\_$aMC@NLO \cite{alwall2014automated} event generator with the center-of-mass energy fixed at 14 TeV. While the SM background events are generated using the inbuilt  SM model file in the MadGraph repository. Parton showers and hadronisation for both signal and background were performed with Pythia 8 \cite{sjostrand2008brief}  followed by detector simulation with DELPHES 3 \cite{de2014delphes}. The subsequent reconstruction of the events followed by a detailed cut-based analysis is presented using MadAnalysis 5 framework \cite{Conte:2012fm}. However the input features for feeding the Neural Networks are reconstructed using CERN ROOT\cite{root} analysis framework. The 2HDM model used for generating the dataset was built using the Feynrules program \cite{alloul2014feynrules, Degrande_2012}. 

For the $p p \rightarrow H^{\pm} \rightarrow W^{\pm}h \rightarrow j j b b$ signal, the dominant background comes from ttbar+jets, t+jets, ZZ+jets, WZ+jets, WW+jets, Z+jets, W+jets processes. To remove the soft jets and soft leptons which arise during hadronization the following identification cuts are imposed on the final states 
\begin{equation}
\begin{split}
    P_{T}^{j} \geq 20 \text{ GeV}, \quad  P_{T}^{l} \geq 10 \text{ GeV}, 
   \\  \abs{\eta^{j}} \leq 5.0, \quad \abs{\eta^{l}} \text{ \& } \abs{\eta^{b}} \leq 2.5
\end{split}
\end{equation}
A wider window of pseudorapidity is chosen so as to not lose too many signal events in the process. Further, to ensure that all the final states are optimally isolated, the following separation cuts are imposed
\begin{equation}
    \Delta R_{jj} = \Delta R_{j l} = \Delta R_{b b} \geq 0.4
\end{equation}

\begin{table*}[htb]
\caption{Number of events after passing the Neural Network output cut.}
\begin{tabular*}{\textwidth}{@{\extracolsep{\fill}}lrrrrrrrrr@{}}
\hline\noalign{\smallskip}
Cut selection & Signal & ttbar+jets & t+jets  & ZZ+jets & WZ+jets &  WW+jets & Z+jets & W+jets & $\frac{S}{\sqrt{S+B}}$ \\ \hline 
Initial & 30000 & 70000 & 50000 & 50000 & 50000 & 50000 & 50000 & 50000  & 47.43 \\
 NN Output $>$ 0.5 & 23441 & 19483 & 11060 & 9858 & 13701 & 11205 & 9278 & 6020 & 72.67 \\
 \hline\noalign{\smallskip}
\end{tabular*}
\end{table*}

With this basic framework in place, the optimization for discovery process of the charged Higgs are proceeded by designing kinematic cuts. Fig. 7 displays the transverse momenta of the leading jet for both the signal corresponding to $M_{H^{+}}=500$ GeV and the SM backgrounds. It is found that a strong cut on the momenta of final state particles is not needed and the signal produces a significant fraction of hard jets with $p_{T} >40$ GeV as one would expect of the particles coming from the decay of charged Higgs boson of mass 500 GeV. Fig. 8 shows the transverse hadronic energy which has higher spread in the signal events. This prompts us to choose $H_{T} > 250$ GeV to eliminate the SM backgrounds.  The invariant mass distribution for the pair of b quarks is displayed in Fig. 9. It shows that there is significant overlap between signal and background making the $M_{b_{1}b_{2}}$ observable unimportant in discriminating signal from background. Finally we turn to the final step in the process of isolating the events in the $M_{b_{1}b_{2}j_{1}j_{2}}$ distribution that corresponds to the decay of charged Higgs boson which is displayed in Fig. 10. In both $M_{b_{1}b_{2}}$ and $M_{b_{1}b_{2}j_{1}j_{2}}$ cases, the signal and SM background overlap to a large extent making the job of isolating  the signal from the backgrounds difficult. 

Finally we present in Table II, the complete cut flow chart that details the impact of each of the kinematic cuts employed on both signal and the various background. The set of kinematic cuts are; 10 $\leq N(j) \leq$ 10, $N(b) \geq$ 3, missing transverse energy (MET) $\geq 20 $ GeV, total transverse hadronic energy $(H_{T}) \geq 250$ GeV. Since a cut on $M_{b_{1}b_{2}}$ and $M_{b_{1}b_{2}j_{1}j_{2}}$ reduces signal by a large extent, it is not applied despite making the discovery prospects higher.  As can be seen in Table II, the progressive cuts have done a good job in systematically suppressing the SM background.  The kinematic cuts reduce the background by 98\% while signal is also lost by 69\%. But the large overlap of the signal and background prompts us to search beyond kinematic cuts based analysis to the era of Machine Learning Techniques.

\subsection{Setup for Machine Learning}
\label{sec:machine_learn}
Here, in this article, a the feed-forward multi-layer artificial Neural Network (ANN) with back-propagation is used. Supervised learning with 50 nodes, 3 hidden layers, and elu activation function for input layers, relu for hidden layers and sigmoid activation function for output layers are used for training for the signal against background. The binary crossentropy (BCE) is used for calculating the loss. Batch normalization and dropout is used to increase the efficiency of the Networks. The input features are the transverse momentum, energy and eta of the jets, b-tagged jets, leptons and the reconstructed di-bjet system.

Fig. 11 and Fig. 13 shows the Neural Network response for the signal against background t+jets and ttbar+jets respectively. The corresponding receiver operating characteristic (ROC) curve are shown in Fig. 12 and Fig. 14. The area under the curve (AUC)  for training aginst t+jets is larger than training against ttbar+jets which shows that the Network response for t+jets is more efficient than ttbar+jets. But the training against t+jets predicts more signal like response for ttbar+jets as shown in Fig. 15. Fig. 16 shows the Neural Network prediction for all the SM backgrounds when trained against ttbar+jets. It is seen that there is clear demarcation between the SM background and the signal. Thus cut is made on the Neural Network output and the number of events passing the cut are shown in Table III along with the significance.  The applied cut reduces the background by 78\% while lossing the signal only by 21\%. 

\section{Conclusions} \label{sec:conclusion}
The discovery of SM-like 125 GeV Higgs at the ATLAS and CMS experiments motivates us  to understand the potential of these experiments to unravel signatures of new physics. The various well motivated extensions of the SM incorporate an enlarged scalar sector with additional charged and neutral Higgs boson. In this article, first various experimental results from Belle and LHC collaborations that put constraints on external parameters of 2HDM Type-II model such as $m_{H^{\pm}}$ and $\tan\beta$ are discussed. Then a collider analysis in the experimentally unexplored region is performed to understand the discovery potential of a charged Higgs boson. The dominant QCD multi-jet background such as ttbar+jets and t+jets were difficult to separate based on kinematic cuts. An advanced method such as Artificial Neural Network training with transverse momentum, energy and eta of the jets, b-tagged jets, and leptons as input features is used. This technique separates signal from background based configuration of the hyperspace parameters. It was found that the kinematic cuts based analysis reduces the background by 98\% while lossing large signal neraly 69\%. Despite the huge constraint on the background, there lies a large overlap of both signal and background region. The applied Machine Learning Technique reduces the signal by 78\% while lossing the signal only by 21\% and there lies distinct separation between the signal and background. This method provides large statistics after cuts for further analysis which will reduce the statistical uncertainties. The Neural Network output cut on 0.5 gave the significance of 72.67 which gives the large prospects of 5$\sigma$ discovery at Collider experiments. 

While the search for charged Higgs is going on at LHC and Belle in full force, it is right time to for signatures that might be hidden from us. This paper summarizes the search strategy that one could employ to discover charged Higgs boson that could be part of the spectrum of a class of models.




\begin{thebibliography}{9}

\bibitem{aad2012observation}
 G.  Aad,  T.  Abajyan,  B.  Abbott, et al., Physics Letters B \textbf{716}, 1 (2012). \\
 DOI: https://doi.org/10.1016/j.physletb.2012.08.020

\bibitem{chatrchyan2012observation}
 S. Chatrchyan, V. Khachatryan, A. M. Sirunyan, et al., Physics Letters B \textbf{716}, 30 (2012). \\
 DOI: https://doi.org/10.1016/j.physletb.2012.08.021
 
\bibitem{hill1993spontaneously}
 C. T. Hill, D. C. Kennedy, T. Onogi, and H.-L. Yu, Physical Review D \textbf{47}, 2940 (1993). \\ 
 DOI: https://doi.org/10.1103/PhysRevD.47.2940
 
\bibitem{Lees_2012}
J.  P.  Lees,  V.  Poireau,  V.  Tisserand, et al.,  Physical Review Letters \textbf{109}, 1079-7114 (2012). \\
DOI: http://dx.doi.org/10.1103/PhysRevLett.109.101802

\bibitem{Huschle_2015} M. Huschle, T. Kuhr, M. Heck, P. Goldenzweig, et al., Physical  Review  D \textbf{92} (2015). \\
DOI: http://dx.doi.org/10.1103/PhysRevD.92.072014

\bibitem{Aaij_2015}  R. Aaij, B. Adeva, M. Adinolfi, A. Affolder, et al., Physical ReviewLetters \textbf{115} (2015). \\
DOI:http://dx.doi.org/10.1103/PhysRevLett.115.111803

\bibitem{Proceedings:2015ona}
Proceedings, 2015 European Physical Society Conference on High Energy Physics, SISSA, Trieste Vol. EPS-HEP (2015) \\
DOI:https://pos.sissa.it/cgi$-$bin/reader/conf.cgi?confid=234

\bibitem{Branco_2012} 
G. Branco, P. Ferreira, L. Lavoura, M. Rebelo, M. Sher,and J. P. Silva, Physics Reports \textbf{516}, 1–102 (2012). \\ DOI:http://dx.doi.org/10.1016/j.physrep.2012.02.002

\bibitem{Moretti_2002}
 S. Moretti, K. Odagiri, P. Richardson, M. H. Seymour,and  B.  R.  Webber,  Journal  of High Energy Physics \textbf{2002}, 028 (2002) \\

 
\bibitem{enberg2015charged}
R. Enberg, W. Klemm, S. Moretti, S. Munir, G. Wouda, Nuclear Physics B \textbf{893} 420-442 (2015). \\
DOI: https://doi.org/10.1016/j.nuclphysb.2015.02.001


\bibitem{misiak2017weak}
M.  Misiak  and  M.  Steinhauser, European  Physical  Journal  C \textbf{77} ,201 (2017).  \\
DOI:http://dx.doi.org/10.1140/epjc/s10052-017-4776-y

\bibitem{hamer2016search}
P. Hamer, A. Frey, A. Abdesselam, I. Adachi, H. Aihara, et al., Physical Review D \textbf{93}, 032007 (2016) \\
DOI: https://doi.org/10.1103/PhysRevD.93.032007

\bibitem{constra}
The ATLAS collaboration., Aad, G., Abbott, B. et al., Journal of high energy physics \textbf{2015}, 206 (2015). \\ https://doi.org/10.1007/JHEP11(2015)206

\bibitem{atkinson}
O. Atkinson, M. Black, A. Lenz, A. Rusov and J. Wynne, Journal of High Energy Physics, 172 \textbf{2022}. \\ 
https://doi.org/10.1007/JHEP04(2022)172

\bibitem{collaborations2013search}
ALEPH Collaboration., DELPHI Collaboration., L3 Collaboration. et al., Eur. Phys. J. C \textbf{73}, 2463 (2013) \\
DOI: https://doi.org/10.1140/epjc/s10052-013-2463-1

\bibitem{akeroyd2017prospects}
 A. Akeroyd, M. Aoki, A. Arhrib, L. Basso, I. Ginzburg, et al., The  European  Physical  Journal  C \textbf{77},  276 (2017). \\
DOI: https://doi.org/10.1140/epjc/s10052-017-4829-2

\bibitem{heavyflavoraveraginggroup2014averages}
H.  F.  A.  Group,  Y.  Amhis,  S.  Banerjee,  et al., Heavy Flavor Averaging Group (HFAG) Collaboration  (2014). arXiv:1412.7515

\bibitem{cms2013search}
 CMS collaboration., Khachatryan, V., Sirunyan, A.M. et al.,   J. High Energ. Phys. \textbf{2015}, 18 (2015). \\ DOI:https://doi.org/10.1007/JHEP11(2015)018 

\bibitem{aad2015search}
G. Aad, B. Abbott, J. Abdallah, S. A. Khalek, R. Aben, et al., Journal of high energy physics \textbf{2015}, 88 (2015) \\
DOI:http://dx.doi.org/10.1007/JHEP03(2015)088

\bibitem{aad2016search}
  G. Aad, B. Abbott, J. Abdallah, R. Aben, M. Abolins, et al.  Journal of high energy physics \textbf{2016}, 127 (2016). \\
  DOI: http://dx.doi.org/10.1007/JHEP03(2016)127
  
\bibitem{binosi2004jaxodraw}
Binosi, D. Theussl, Lukas, Computer Physics Communications, \textbf{163} 76-86 (2004) \\
DOI: http://dx.doi.org/10.1016/j.cpc.2004.05.001


\bibitem{alwall2014automated}
J. Alwall, R. Frederix, S. Frixione, V. Hirschi, F. Maltoni, et al.,  Journal of High Energy Physics \textbf{2014}, 79 (2014). \\
http://dx.doi.org/10.1007/JHEP07(2014)079

\bibitem{sjostrand2008brief}
T. Sjöstrand, S. Mrenna, and P. Skands, Computer  Physics  Communications \textbf{178}, 852 (2008). \\
DOI:https://doi.org/10.1016/j.cpc.2008.01.036

\bibitem{de2014delphes}
 J. De Favereau,  C. Delaere,  P. Demin,  A. Giammanco, et al.,  Journal of High Energy Physics \textbf{2014}, 57 (2014). \\ 
DOI: http://dx.doi.org/10.1007/JHEP02(2014)057

\bibitem{Conte:2012fm}
E. Conte, B. Fuks, G. Serret, Computer Physics Communications, \textbf{184} 1 2013. \\
DOI:http://dx.doi.org/10.1016/j.cpc.2012.09.009

\bibitem{root}
Rene Brun and Fons Rademakers, Nucl. Inst. \& Meth. in Phys. Res. A \textbf{389} 81-86 (1997). \\
DOI: https://zenodo.org/record/3895860$\#.$YOri9y1h2t9

\bibitem{alloul2014feynrules}
 A.  Alloul,  N.  D.  Christensen,  C.  Degrande,  C.  Duhr,and  B.  Fuks, Computer Physics Communications \textbf{185}, 2250 (2014). \\
 DOI:https://doi.org/10.1016/j.cpc.2014.04.012
 
 \bibitem{Degrande_2012}
  C. Degrande, C. Duhr, B. Fuks, D. Grellscheid, O. Mat-telaer, and T. Reiter, Computer Physics Communications \textbf{183}, 1201–1214 (2012) \\
DOI:http://dx.doi.org/10.1016/j.cpc.2012.01.022

\end{thebibliography}
\end{document}